\documentclass[runningheads]{llncs}
\usepackage{graphicx}
\usepackage{wrapfig}
\usepackage{comment}

\begin{document}
\title{PMSD: Data-Driven Simulation Using System Dynamics and Process Mining  \thanks{\scriptsize{ Funded by the Deutsche Forschungsgemeinschaft (DFG, German Research Foundation) under Germany's Excellence Strategy – EXC 2023 Internet of Production- Project ID: 390621612. We also thank the Alexander von Humboldt (AvH) Stiftung for supporting our research.}}}

\author{Mahsa Pourbafrani \and
Wil M. P. van der Aalst}
\authorrunning{M. Pourbafrani and Wil M. P. van der Aalst}
\institute{Chair of Process and Data Science, RWTH Aachen University, Germany \\
 \email{\{mahsa.bafrani,wvdaalst\}@pads.rwth-aachen.de}
 }
\titlerunning{PMSD}
\maketitle
\setcounter{footnote}{0}      
\begin{abstract}
Process mining extends far beyond process discovery and conformance checking, and also provides techniques for bottleneck analysis and organizational mining. However, 
these techniques are mostly \emph{backward-looking}. \emph{PMSD} is a web application tool that supports forward-looking simulation techniques. It transforms the event data and process mining results into a simulation model which can be executed and validated. \emph{PMSD} includes log transformation, time window selection, relation detection, interactive model generation, simulating and validating the models in the form of system dynamics, i.e., a technique for an aggregated simulation. The results of the modules are visualized in the tool for a better interpretation.

\end{abstract}

\section{Introduction}
Process mining uses stored event data of organizations, i.e., event logs, to provide actionable insights for organizations~\cite{DBLP:books/sp/Aalst16}.
Different tools address process discovery, performance analysis, bottleneck analysis, and deviation detection.
Yet, the gap between the \emph{backward-looking} and the \emph{forward-looking} process mining techniques remains.
Traditional forward-looking techniques as mentioned in \cite{SummerSimKeynote2018}, use events in the process as a basis of simulation. They aimed to mimic the process at the level of detail and simulate it. In more recent simulation tool such as \cite{DBLP:conf/bpm/CamargoDR19}, different level of detail for simulation is acquired, e,g., duration of activities and the flow of activities are used.  
\begin{figure}[bt]
    \centering
    \includegraphics[width=0.78\textwidth, height = 0.11\textheight]{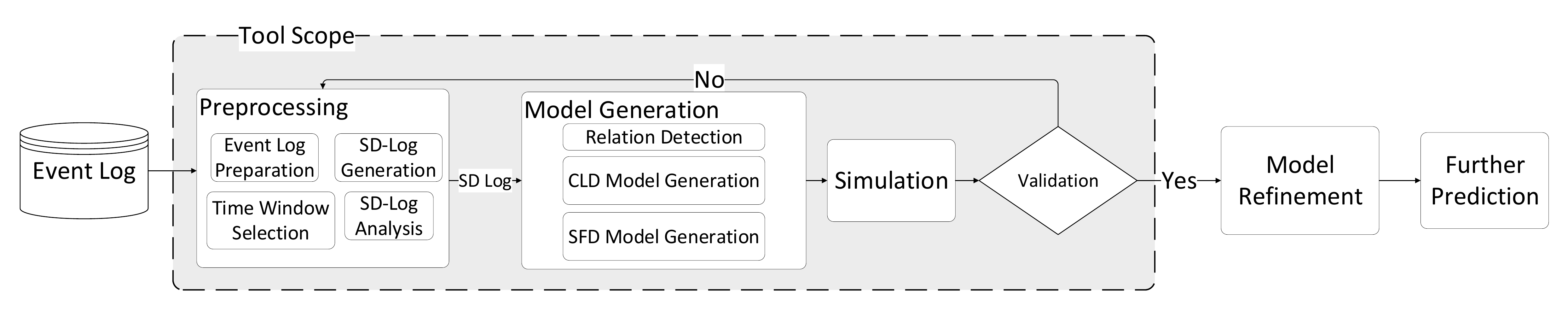}
    \caption{Our proposed framework for using process mining and system dynamics together in order to design valid models to support scenario-based prediction of business processes in \cite{pourbafrani2019scenario}. This paper focuses on the \emph{developed tool}, i.e., the highlighted step.}
    \label{fig:overview}
\end{figure}
Moreover, the \emph{Monte Carlo} technique is used in the \emph{pm4py} tool\footnote{\scriptsize{ http://pm4py.pads.rwth-aachen.de}} for simulating discovered Petri nets.

\begin{wrapfigure}{l}{0.38\textwidth}
\centering
    \includegraphics[width=0.3\textwidth, height = 0.14\textheight]{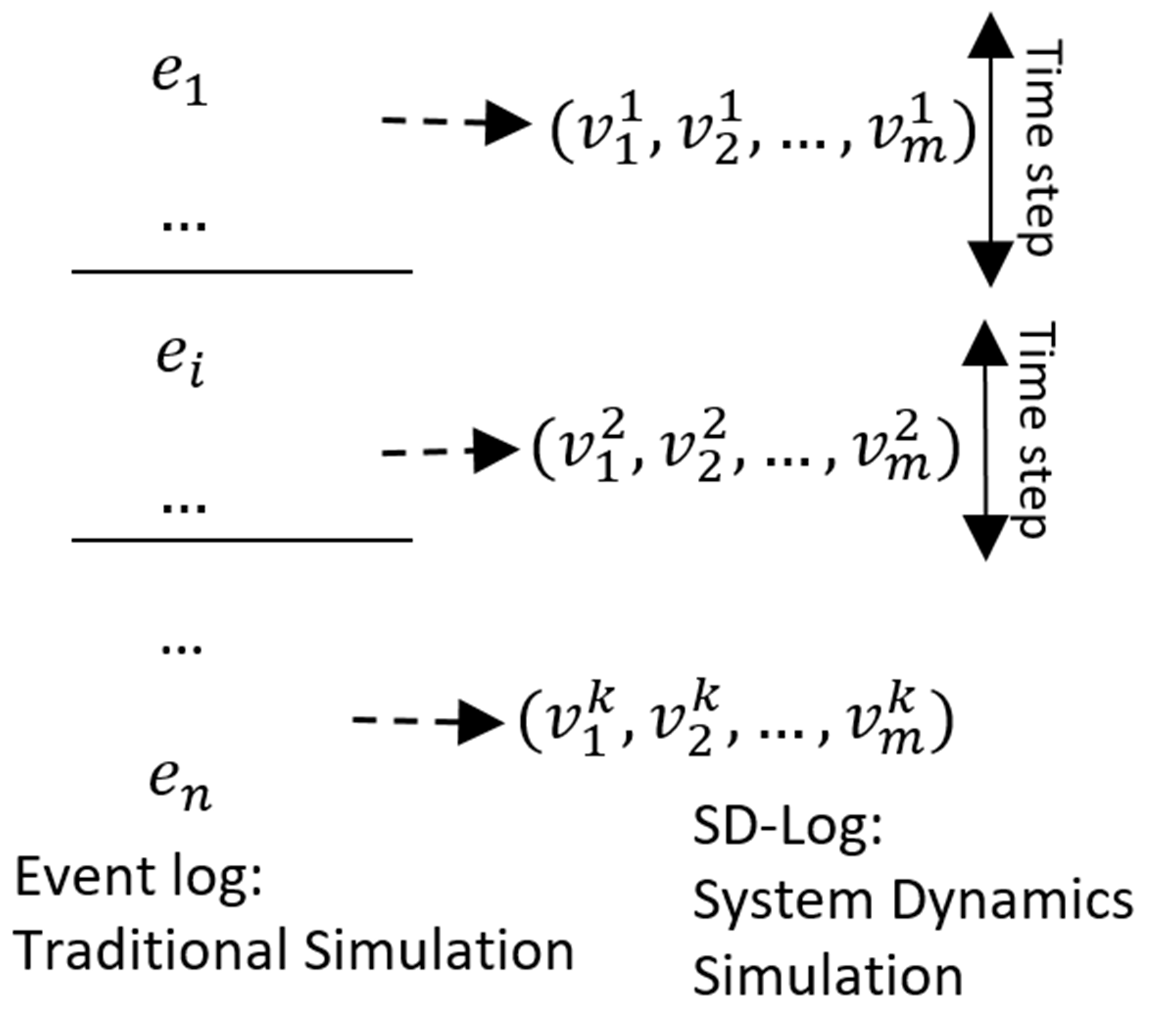}
    \caption{Traditional Simulation vs. \emph{PMSD}. We extract possible variables (\emph{m}) over time steps (\emph{k}).}
    \label{fig:desvssd}
\end{wrapfigure}

In \emph{PMSD}, we use the idea that a simulation model can be learned from the event data at an aggregated level. The traditional connections between process mining and simulation mainly use a descriptive model discovered in the discovery step to enrich the simulation models at the level of the process instances, e.g., Discrete Event Simulation (DES).
The presented tool is the result of our approach in generating simulation results for business processes at an aggregated level providing the option to add external factors into the simulation \cite{pourbafrani2019scenario}.
Figure~\ref{fig:overview} shows the overview of the approach starting from an event log and ending with a scenario-based simulation model. The steps indicated in the highlighted parts are supported by the tool. We extract possible variables from the process in different steps of time instead of taking the events into account for the simulation as shown in Fig.~\ref{fig:desvssd}.

The Model generation module is introduced in \cite{mahsaBIS2020} and the preprocessing step is presented in \cite{mahsaTimeseries}.
The event log is transformed into a set of variables over time and the values of these variables form the System Dynamics logs (SD-Logs). To generate more stable SD-Logs, we use time series analysis over the values. The relations between variables over time in the SD-Log are used for creating the system dynamics models. We support both causal loop diagrams (CLD) and stock-flow diagrams (SFD). 
System dynamics models the systems and their relations with the environment~\cite{sterman2002system}. 
CLDs represent these conceptual relationships and SFDs model the underlying equations using \emph{stock}, \emph{flow} and \emph{variable} notations. Flows add/remove to/from the values of stocks, also, variables affect/get affected by the flows, other variables.
\emph{PMSD} provides insights through the processes over time which can be hidden from the user, e.g., a nonlinear relation between the workload of resources and the speed of performing tasks. 

\section{Description of Functionalities}
In our approach, the possible process variables are extracted over time, e.g., arrival rate per day and average service time per day. The newly generated log (SD-Log) is the cornerstone of the simulation. 
The preprocessing step and extracting the best parameters in the framework by means of time series analysis proposed in \cite{mahsaTimeseries}.
To form a valid system dynamics model, we have to discover all the relations, i.e., linear and nonlinear correlations, between the generated process variables over time as introduced in \cite{mahsaBIS2020}. 
Analyzing a process and creating aggregated features of the process over time (process variables) for further analyses is the main focus of the tool.

\begin{figure}[bt]
    \centering
    \includegraphics[width=0.77\textwidth, height = 0.145\textheight]{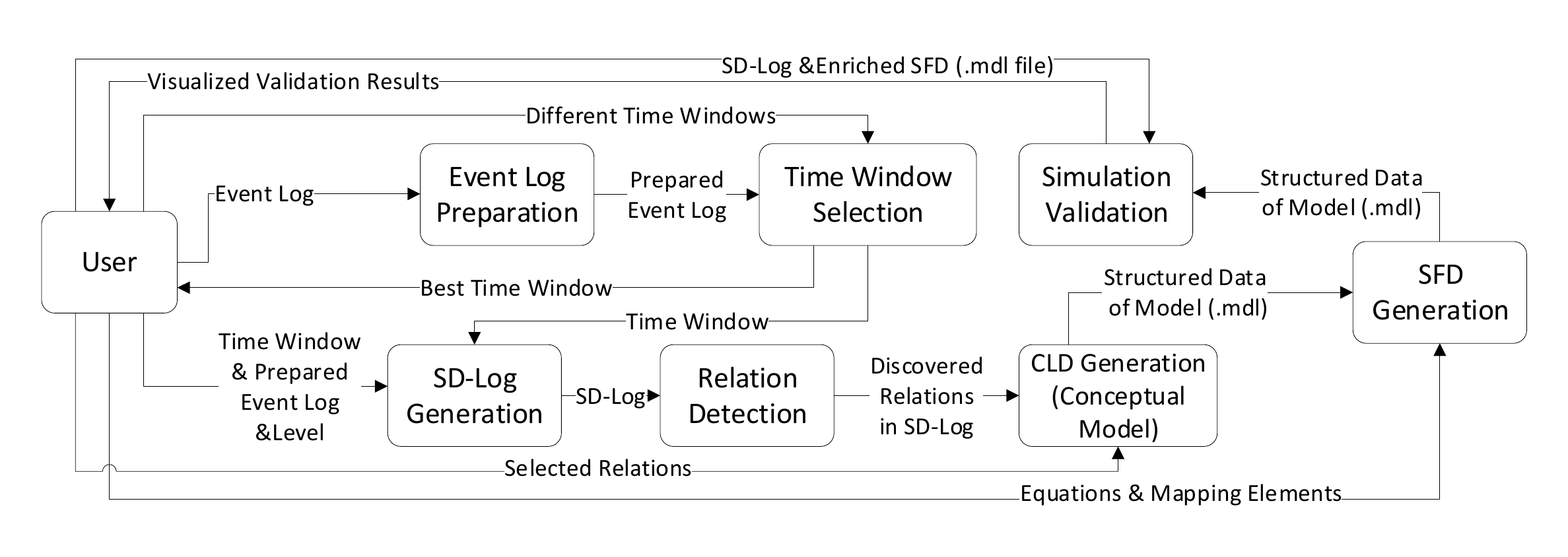}
    \caption{Data flow diagram of the \emph{PMSD} including data flow between the user and the main modules as well as the background flow of data between the modules.}  
    \label{fig:appdatamodel}
\end{figure}

\emph{PMSD} is being designed in such a way that in all the steps, the outputs are accessible for users. 
Figure~\ref{fig:appdatamodel} depicts the data flow diagram of the application. 
The inputs and generated outputs in each module and the interactions with the user are shown. 
The generated SD-Logs including active steps in the processes as well as all the steps for the different selected time windows in the form of \emph{.csv} are captured. Also, all the designed CLDs and SFDs in the \emph{.mdl} format are stored locally for the user. 
To run the tool locally, the home page can be accessed via any browser using the \emph{http://127.0.0.1:5000} URL. All the modules are designed as different tabs and are visually accessible. 
\emph{PMSD} is a fully interactive tool with a user interface based on \emph{Python} and \emph{Flask} technology. The results of the steps are shown graphically to provide an easier interpretation possible. It contains 8 tabs and each tab can be run separately with different inputs/output of the other modules/tabs.
Currently, the following components are available:
\begin{itemize}
  \item \emph{Event log transformation} indicates the main attributes of the event log, discovers the directly follows graph, and presents the event log's information. 
   \item \emph{Time window selection} assesses the quality of the user's preference for selecting a time window for generating simulation data. 
  \item \emph{Simulation log generator} uses the transformed event log and the selected time window to generate simulation data (SD-Log). It generates an SD-Log for different aspects and levels, i.e., general process, organizational, and activity aspects.
  For instance, an SD-Log of the general aspect of a process includes the arrival rate of the process, and average service time in the process and other possible measurable variables per day. 
  \item \emph{Relation detection} investigates whether there is any strong relationship between the variables in the extracted SD-Log. 
  Furthermore, the user can look for the relations between variables in different steps of time.
  \item \emph{Detailed relations}, presents the existing relations between every two variables in the SD-log for further investigation on the types of relations. 
  \item \emph{Interactive conceptual model generation} provides the option for the user to choose between all the strong relations discovered in the relation detection module and creates CLD, i.e., effects and relations between process variables. It generates both the graphical model in the tool and the \emph{.mdl} (text format) file to be used in most of the system dynamics tools, e.g., \emph{Vensim}\footnote{\scriptsize{ www.vensim.com}}. 
  \item \emph{Interactive stock-flow diagram} generates SFDs graphically in \emph{PMSD} and the (.mdl) file. The relations are directly transformed from the CLD (previous step) and the user can map the process variables to the SFD elements. 
  \item \emph{Simulation and validation} simulates the SFD model using the values in the SD-Log and validates the results using the pair-wise comparison of the SD-Log and simulation results values and their distributions. 
\end{itemize}

\section{Maturity of the Tool}
The evaluation results of our proposed forward-looking approaches in process mining are represented using different modules of the tool. 
\emph{PMSD} along with a tutorial and a screen-cast is available on GitHub.\footnote{\scriptsize{https://github.com/mbafrani/PMSD}} It has also been used in some industrial projects, e.g., in the project of \emph{Internet of Production} in the context of \emph{Industry 4.0}.
In \cite{DBLP:conf/ihsi/PourbafraniZA20}, part of the results of using \emph{PMSD} for the production line is presented. By an example, i.e., an event log of a call center designed by the CPN tool, we show some similar results.

\begin{wrapfigure}{l}{0.5\textwidth}
\vspace{-10 mm}
  \begin{center}
    \includegraphics[width=0.5\textwidth, height= 0.2\textheight]{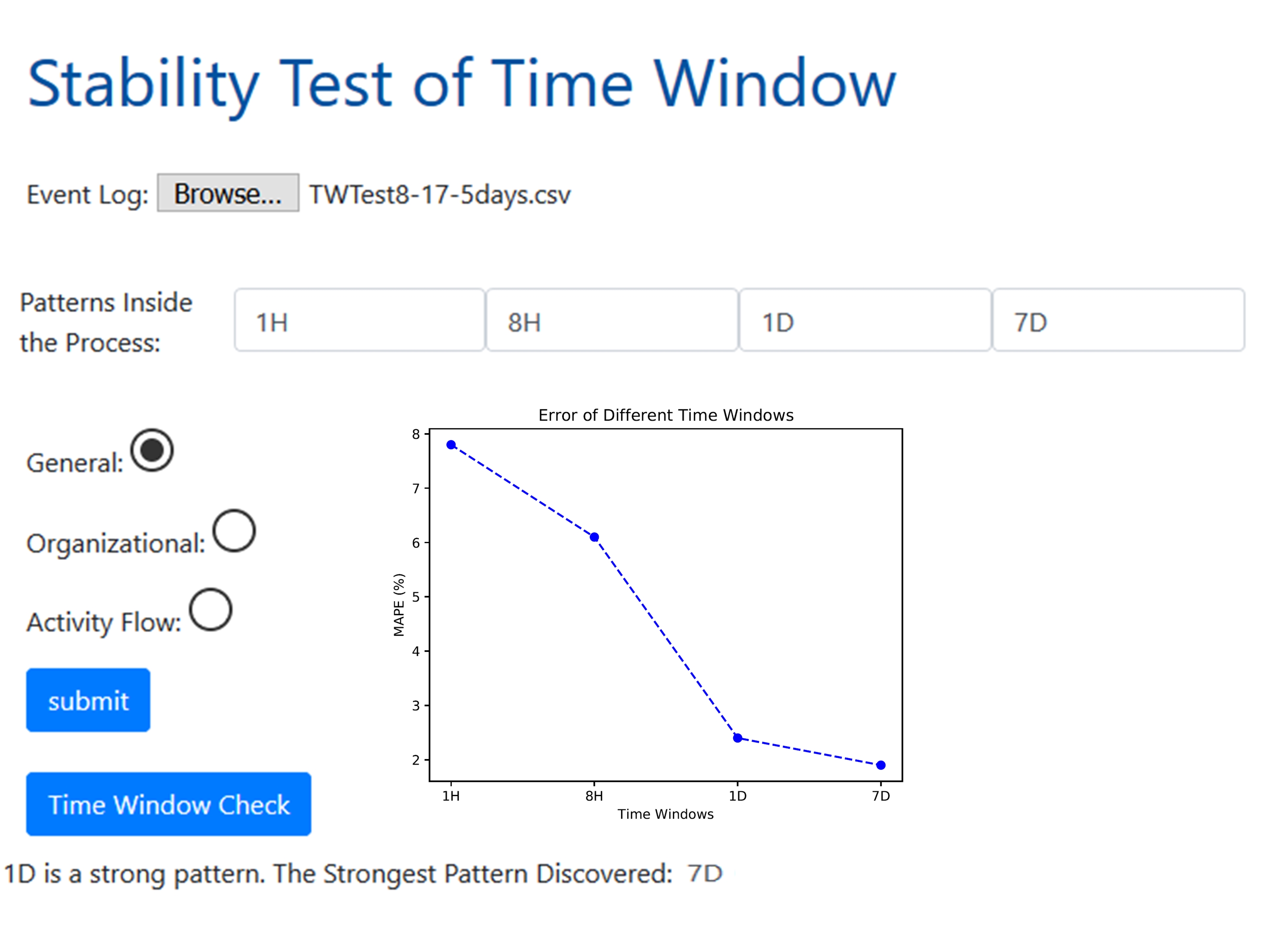}
  \end{center}
  \caption{\emph{Stability test} showing the error of training models for the time windows.}
  \label{fig:stabiltiy test}
  \vspace{-7 mm}
\end{wrapfigure}
\begin{figure}[bt]
\centering
    \includegraphics[width=0.75\textwidth,height=0.14\textheight]{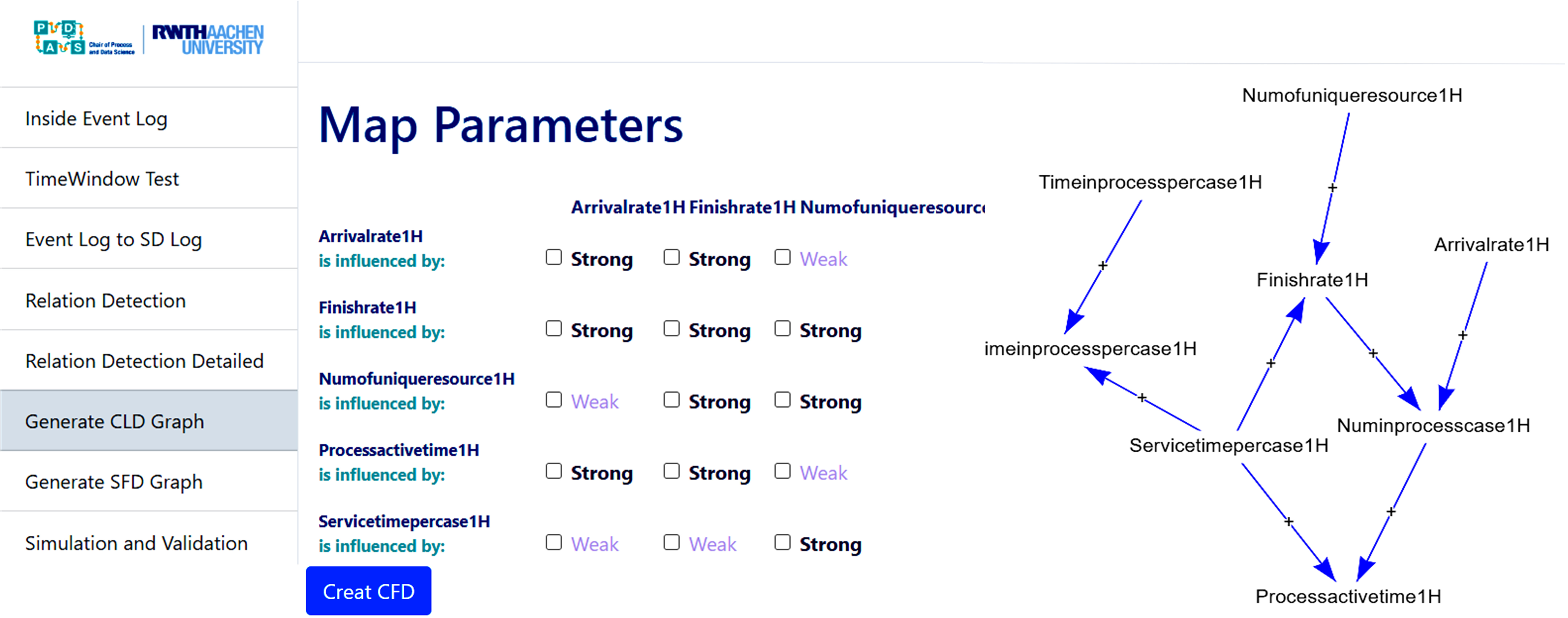}
    \caption{The \emph{conceptual modeling} section showing the detected relations and their strength between the variables. The user is able to select among the selected relations.}
    \label{fig:cLDGenerationPage}
\end{figure}

We use different suggested time windows to extract values over time for the possible process variables using the time window test. The result in Fig. \ref{fig:stabiltiy test} shows the selected time windows by the user and the errors of trained models for each time window. 
Figure \ref{fig:cLDGenerationPage} represents the user interface for selecting the strong detected relations between the variables.
Finally, by uploading the generated SFD and SD-log (both are automatically generated), the automatic simulation is performed and the validation results are shown in validation module.
The results include a comparison between the real values and the simulated ones and their distributions for the selected variables. 

\section{Conclusion}
In this paper, we introduced \emph{PMSD} to support designing system dynamics models for simulation in the context of business processes.
Using \emph{PMSD}, we look into the processes at different aggregation levels, e.g., hourly or daily, as well as different aspects, e.g., overall process or organizational aspects. 
The provided user interface and the graphical outputs make the interpretation of the results easy.
Applying \emph{PMSD}, the underlying effects and relations at the instance level can be detected and modeled in an aggregated manner.
Besides the option to simulate and validate the models directly in the tool, the models can be simulated or refined by adding external variables using simulation software like \emph{Vensim}. 
\bibliographystyle{splncs04}
\bibliography{Reference}

\end{document}